\documentstyle[aps,preprint]{revtex}
\def\abs#1{\left|{#1}\right|}
\let \nn = \nonumber

\input psfig
\begin{document}
\draft
\preprint{OHSTPY-HEP-T-95-028}
\date{December 27, 1995}
\title{ GUT Scale Threshold Corrections \\in a Complete Supersymmetric SO(10) Model:\\
$\alpha_s(M_Z)$  vs. Proton Lifetime}

\author{Vincent Lucas$^{1}$ and Stuart Raby$^{1,2}$}
\address{
$^{1}$Department of Physics, The Ohio State University, 174 W. 18th Ave., Columbus, OH 
43210}
\address{
$^{2}$raby@mps.ohio-state.edu}

\maketitle

\begin{abstract}

We show that one loop GUT scale threshold corrections to gauge couplings are a significant 
constraint on the GUT symmetry breaking sector of the theory.   The one loop threshold
corrections relate the prediction for $\alpha_s(M_Z)$ to the proton lifetime.   We have
calculated these corrections in a new complete SO(10) SUSY GUT.  The results are consistent
with the low energy measurement of $\alpha_s(M_Z)$.  We have also calculated the proton
lifetime and branching ratios in this model.  We show that proton decay rates provide a
powerful test for theories of fermion masses.   \end{abstract}

\pacs{PACS numbers: 12.15.Ff, 12.15.Hh, 12.60.Jv }

\narrowtext

\section{Introduction}
It has been known for some time now that Supersymmetric [SUSY] Grand Unified Theories 
[GUTs]\cite{DRW}

$\bullet$ suppress the non-SUSY GUT prediction for proton decay, via heavy gauge boson 
exchange, by increasing the GUT scale, and 

$\bullet$ lead to a 10\% increase (for one pair of Higgs doublets) in the GUT prediction 
for the weak mixing angle, $\sin^2\theta_W$, compared to non-SUSY GUTs. 

 Now, since $\sin^2\theta_W$ is known to better than 1\% accuracy, one naturally uses the 
values of the two low energy parameters, the fine structure constant, $\alpha$, and
$\sin^2\theta_W$ as input to determine the GUT coupling, $\alpha_G$, and the GUT scale,
$M_G$, and predicts the value of the strong coupling, $\alpha_s$.  [Note, all low energy
parameters are typically evaluated at the scale, $M_Z$.] This prediction is in excellent
agreement with the low energy data\cite{webber,el,lp}.  The global fit to $\alpha_s$ from
measurements at all energies\cite{webber} gives $\alpha_s = 0.117 \pm 0.005$ with low (high)
energy measurements favoring values of $\alpha_s \le (\ge)\; 0.12$.  A global fit to all
electroweak data\cite{el} gives $\alpha_s = 0.127 \pm 0.005 \pm 0.002 \pm 0.001$.  In
comparison, SUSY GUTs predict\cite{lp} $\alpha_s = 0.127 \pm 0.005 \pm 0.002$, where the
first error is the uncertainty in the low energy sparticle spectrum, and the second, the top
and Higgs masses.  Note that the non-supersymmetric GUT prediction\cite{langacker} $\alpha_s =
0.073 \pm 0.001 \pm 0.001$ is clearly ruled out.
 
In addition, SUSY GUTs contain new sources of proton decay (dimension 5 operators) mediated by 
color triplet Higgs fermions\cite{wsy,PDECAY}.  They typically lead to the dominant decay
mode $p \rightarrow K^+ \bar{\nu}$ with the lifetime, $\tau_p \propto \tilde{M}_t^2$, where
$\tilde{M}_t$ is an effective color triplet mass.
 
These results are valid for any grand unified symmetry group, as long as
SU(3)$\times$ SU(2)$\times$ U(1) is embedded in an unbroken simple subgroup above the
GUT scale.  Consideration of fermion masses, however, leads us to the special grand unified
symmetry group SO(10)\cite{SO(10)}.  It is the unique group which combines all the fermions
of one family \{u, d, e, $\nu$\} into one irreducible 16 dimensional representation with the
addition of only one electroweak singlet state, $\bar{\nu}$, the so-called right-handed
neutrino.  Consequently, SO(10) Clebschs can be used to relate all charged fermion mass
matrices.  In this context, a paper by Anderson et al.\cite{ADHRS}[Paper I] showed how all
charged fermion masses and mixing angles can be reasonably described in terms of four
dominant effective mass operators at $M_G$.  Hall and one of us (S.R.)\cite{hr} [Paper II]
then showed how to extend the effective theory at $M_G$ to a complete SO(10) SUSY GUT valid
up to the Planck or string scales, $M$.  That theory included several sectors necessary for
-- (1) GUT symmetry breaking,  (2) doublet-triplet splitting (Higgs sector), 
(3) charged fermion masses (fermion mass sector) and also 
(4) giving a large mass to the electroweak singlet neutrinos (neutrino mass sector).  It
contained sufficient symmetry to be ``natural."  Hence, it included all operators consistent
with the symmetries and only contained the SO(10) states\footnote{These two features are
consistent with a stringy origin to such a model\cite{CCL}.} ---  \begin{equation}(n_{16} +
3)\;{\bf 16} + n_{16} \; \bar{\bf 16} + n_{10} \; {\bf 10} + n_{45}  \; {\bf 45} + n_{54} \;
{\bf 54} + n_1 \; {\bf 1}  \label{eq:states} \end{equation} with $n_{16}, n_{10}, n_{45},
n_{54}, n_1$ specific integers.  Finally, the  effective mass operators of paper I  were
recovered upon integrating out states with mass greater than $M_G$.

In any complete GUT, the prediction for $\alpha_s$ receives corrections at one loop from
thresholds at the weak scale, including the sparticle spectrum, and at the GUT scale from states
with mass of order $M_G$.  The dominant contribution at $M_G$ comes from states in the symmetry
breaking and Higgs sectors.

The Higgs sector, on the other hand, also affects the proton decay rate.  The effective color
triplet Higgs  fermion mass, $\tilde{M}_t$, must be significantly larger than $M_G$ in order to
increase the proton lifetime. The Higgs doublets on the other hand must remain massless at
the GUT scale.  

In this paper,

$\bullet$ we show that the threshold corrections due to doublet-triplet splitting, in the 
simplest models, increases the predicted value for $\alpha_s(M_Z)$.  Thus, while the GUT
predictions for $\alpha_s$ and $\tau_p$ are unrelated at the tree level,  at one loop the two 
effects are coupled.  One cannot suppress proton decay without at the same moment increasing
the prediction for $\alpha_s$.  We then show that the model of paper II is ruled out by this
effect.

$\bullet$ We also present a new complete SUSY GUT which appears to be
consistent with all low energy data at the $1 \sigma$ level\cite{bcrw}.  This model has in fact a 
much simpler GUT symmetry breaking sector than that in paper II with fewer states.   We present
some typical results for proton decay rates in this model.  In a future paper\cite{lr}, we will
present a more detailed study.

\section{One loop threshold corrections at $M_G$.}

One loop threshold corrections at $M_G$ for gauge couplings are given by\cite{threshold}
\begin{equation}
\alpha_i^{-1}(M_G)=\alpha_{G}^{-1}-\Delta_i
\end{equation}
where $\Delta_i$ is the leading log threshold correction to $\alpha_i$. Furthermore,
\begin{equation}
\label{eq:1}
\Delta_i={1 \over 2 \pi} \sum_\zeta b_i^\zeta \log \abs{M_\zeta \over M_G}
\end{equation}
where the sum is over all superheavy particles and $b_i^\zeta$ is the contribution the 
superheavy particle $\zeta$ would make to the beta function coefficient $b_i$ if the particle
were not integrated out at $M_G$.

At one loop the definition of the GUT scale is somewhat arbitrary.  In order to avoid large 
logarithms it must certainly be at the geometric mean of the heavy masses; otherwise we are free
to choose its value.  A particularly convenient choice is to define $M_G$ as the scale 
where the two gauge couplings, $\alpha_i, \;\; i = 1,2$, meet.  At this point $\Delta_1 =
\Delta_2$ and we define \begin{equation}\tilde\alpha_G \equiv \alpha_1(M_G) =
\alpha_2(M_G).\end{equation} Then define \begin{equation}\epsilon_3 \equiv  (\alpha_3(M_G)
- \tilde\alpha_G)/\tilde\alpha_G, \end{equation}
i.e. the relative shift in $\alpha_3$ at $M_G$. In general, a value of $\epsilon_3 \sim - (2 - 3
\%)$ is needed to obtain $\alpha_s \sim \; 0.12$.

\subsection{Formula for $\epsilon_3$ in a general SO(10) theory}

The most general SO(10) SUSY GUT we consider includes the complete multiplets listed in 
equation (\ref{eq:states}) with arbitrary values of 
$n_{16}, n_{10}, n_{45}, n_{54}, n_1$.  These states can be decomposed into their 
SU(3) $\times$ SU(2) $\times$ U(1) content by first considering the decomposition to SU(5)
[see table ~\ref{t:su5decomp}] and then the rest of the way \cite{slansky}. 

We have derived a general formula for $\epsilon_3$ in this case (see appendix 1) given by the
expression (valid
 to lowest order in $\tilde\alpha_G$)
\begin{eqnarray}
\epsilon_3 & = &  {\tilde\alpha_G \over 2 \pi}  \sum_\gamma[(b_3^\gamma
- b_2^\gamma)- {1 \over 2}(b_2^\gamma - b_1^\gamma)] \log \abs{{\det}' {M_\gamma \over M_G}} \label{eq:epsilon}
\end{eqnarray}
where 
\begin{equation}
{\det}' M_\gamma = \left\{  
\matrix{\det M_\gamma \hfill & {\rm if} & \gamma=t,g,w,s,\sigma \hfill \cr
\det M_\gamma \over {(M_{gaugino}^\gamma)^4} \hfill & 
{\rm if} & \gamma=q,u,e,x \hfill \cr
\det M'_d & {\rm if} & \gamma = d \cr} 
\right.   \label{eq:effdet}
\end{equation}
where  $M'_d$ is defined as the doublet mass matrix $M_d$ with the massless Higgs doublets 
projected out.\footnote{Our notation for the states discussed above is found in table
~\ref{t:states}.}  Note, $\epsilon_3$ only depends on the number of states in the theory 
through the mass matrices  $M_\gamma$ in each charge sector $\gamma$.   Moreover, the
effective determinants, defined above, explicitly take into account the special contributions
of vector multiplets to the threshold corrections.  

Putting in the values of $b_i^\gamma$ for $i = 1,2,3$ into
eqn. (\ref{eq:epsilon}) we find

$$\epsilon_3={\tilde\alpha_G \over \pi} \biggl( {3\over 2}\log\abs{{\det}'{M_g \over M_G}} 
- {3\over2}\log\abs{{\det}'{M_w \over M_G}} \biggr.$$
$$  +{33\over10}\log\abs{{\det}'{M_s \over M_G}}
-{21\over10}\log\abs{{\det}'{M_\sigma \over M_G}}  +
  {9\over10}\log\abs{{\det}'{M_u \over M_G}}
+{3 \over 10}\log\abs{{\det}'{M_e \over M_G}} -
  {6\over5}\log\abs{{\det}'{M_q \over M_G}}$$
\begin{equation}
\biggl.
	-{3 \over 5}\log\abs{{\det}'{M_d \over M_G}} +
  {3\over5}\log\abs{{\det}'{M_t \over M_G}} \biggr) .  \label{eq:grandexpression}
\end{equation}

The dominant contribution to  $\epsilon_3$ comes from the GUT symmetry breaking and electroweak
Higgs sectors with superspace potential $W_{sym\, breaking}$ and $W_{Higgs}$, respectively. 
An additional small contribution comes from the fermion mass sector.  These sectors are by
reasons of  ``naturalness" necessarily invariant under several U(1) symmetries (including an R
symmetry).  In appendix 2 we show that these symmetries impose stringent constraints on the
form of  $\epsilon_3$.   In brief,  $\epsilon_3$ is 
implicitly a function of the vacuum expectation values [vevs] of fields in the symmetry
breaking sector.  These vevs transform under the U(1) and R symmetries.  As a
consequence of the invariance, we find 
\begin{equation} {\epsilon}_3=f(\zeta_1,\ldots,\zeta_m) +
{3 \tilde\alpha_G \over 5\pi}\log \abs{{{\det}\bar M_t  \over M_G {\det} \bar M'_d}} +
\cdots \label{eq:ep3} \end{equation}  
where the first term represents the contributions from 
$W_{sym\, breaking}$. It is only a function of U(1) and R invariant products of powers of vevs
\{$\zeta_i$\}.  The second term, coming from the Higgs sector, is discussed further in
the next section and the ellipsis refers to the small additional contribution arising from the
fermion mass sector.   Note, $\bar M_t,\; \bar M'_d$ only include those states, from 
$5$s and $\bar 5$s of SU(5) contained in $W_{sym\, breaking}$ and $W_{Higgs}$, {\em which mix
with the Higgs sector} (see the next section).

\subsection{The dependence of $\epsilon_3$ on the Higgs sector}
We assume the Higgs sector includes any number of {\bf 10}s with, for simplicity, only one of 
these, say ${\bf 10_1}$, coupling to light fermions. We also include the doublet-triplet
splitting mechanism introduced by Dimopoulos-Wilczek~\cite{DW}. Accordingly, the terms in the superspace potential relevant for doublet -
triplet splitting are of the form
\begin{equation} W_{d-t}= \bar{5}_2\; [a_1 \,{3 \over 2}
\;(B~-~L)] \;5_1 - \bar{5}_1 \;[a_1 \,{3 \over 2}\;(B~-~L)]\; 5_2+ \sum_{i, j \ge 2}M_{ij}
\;\bar{5}_i\; 5_j \label{eq:Wdt} \end{equation} where  $a_1 \,{3 \over 2}\; (B~-~L)$ is the
vev of the field $A_1$ in the 45 dimensional representation and  $(B~-~L)$ is the [Baryon -
Lepton number] charge matrix (see eqn. (\ref{eq:a1})).   Thus, since the doublets in $5_1,
\bar{5}_1 \subset 10_1$ have zero $B~-~L$, they remain massless.  [Note, some of the 
$\bar{5}_i \,,5_j$ states may come from $16$ and $\overline{16}$ representations,
respectively, 
 in $W_{sym\, breaking}$. We include, however, only those states which mix with $10_1$.] With
this superpotential, we have $ \bar M'_d \equiv M[d]$, and the triplet mass matrix, $\bar
M_t$, with non-vanishing determinant, includes the terms which mix states in $10_1$ with 
$\bar{5}_i ,\;5_j,\; {\rm for} \;i,j \ge 2$ and the sub-matrix $M[t]$, where the matrix
$M[d],\; (M[t])$ is given by $M$ with Clebschs appropriate for doublets (triplets).  

Proton decay in this model is
mediated by the Higgses in $10_1$.  Hence, the coefficient of the resulting effective dimension 5 
baryon violating operators\cite{wsy,PDECAY} is given by the inverse of the triplet mass
matrix in the 11 direction, i.e. $$(M^{-1}_t)_{11} \equiv {{\det} M[t] \over {\det} \bar
M_t} = {{\det} \bar M'_d \over {\det} \bar M_t} \times  g $$ where $g \equiv { {\det} M[t]
\over {\det} M[d]}$.  We show, in appendix 2, that  $ g = g(\zeta_1, \cdots, \zeta_m) $
is a holomorphic function of the set of U(1) and R invariant products of powers of vevs.  If we now
define the effective triplet mass by $\tilde{M_t}^{-1} \equiv (M^{-1}_t)_{11}$,
we obtain the final form for the factor in eqn. (\ref{eq:ep3}) --  ${{\det} \bar M_t \over {\det} \bar M'_d} = \tilde M_t \; g(\zeta_1, \cdots,
\zeta_m)$. As a consequence, eqn. ~(\ref{eq:ep3}) becomes \begin{equation}
{\epsilon}_3=F(\zeta_1,\ldots,\zeta_m) + {3 \tilde\alpha_G \over 5\pi}\log \abs{{
\tilde{M_t} \over M_G}} + \cdots \end{equation} where $f$ and $g$ are absorbed
in the function $F$.  We thus find that suppressing the proton decay rate by increasing the
ratio ${\tilde{M_t}  \over M_G}$ has the effect of increasing the value of $\epsilon_3$ and
consequently increasing the value of $\alpha_s$.\footnote{This observation in the
context of minimal SU(5), where in that case $\tilde M_t$ is the color triplet Higgs mass,
was discussed earlier by Hisano et al.\cite{PDECAY}.}

\subsection{An Example : $\epsilon_3$ in the model of paper II}
As an example, for the particular model of paper II, there is only one independent invariant
ratio of the GUT scale vevs given by $\zeta= {{v \overline v}\over a_1 a_2}$.  In principle,
$\epsilon_3$ could depend on an arbitrary function $f(\zeta)$.  However, when we put the
effective determinants for this model in equation ~(\ref{eq:grandexpression}) we find
\begin{equation} \epsilon_3= {3\, \tilde\alpha_G \over 5 \pi} \biggl\{21 \,\log(2) + 
\log \abs{{\tilde{M_t} \over M_G}}\biggr\}. \label{eq:modelII} \end{equation} 
Remarkably, all dependence of $\epsilon_3$ on
the GUT scale vevs has dropped out except for the dependence on $\tilde{M_t}$. Unfortunately, the
large positive constant that appears in the expression means that in order to 
get $\epsilon_3$ negative, the  proton's lifetime would be many orders of magnitude lower
than the experimental bound. In addition, changing the Yukawa coupling coefficients of the
terms of $W_{sym\, breaking}$ cannot cure this problem. For the most part, changing the
Yukawa coupling constants of one of the terms of $W_{sym\, breaking}$ has the result of
multiplying the effective determinants of the mass matrices $M_{\gamma}$
in eqn. (\ref{eq:epsilon}), for states $\gamma$ in a complete SU(5) multiplet, all by the
same amount. Thus, this multiplicative factor has no effect on $\epsilon_3$. Hence, as a
consequence of (\ref{eq:modelII}), we find that the symmetry breaking sector of the model
of paper II is ruled out by  the dual constraints coming from the low energy measurement of
$\alpha_s$ and the proton lifetime.  

Is it possible to find a symmetry breaking sector which has
all the U(1) symmetries required for the naturalness of the theory and is consistent with the
constraints of $\alpha_s$ and $\tau_p$?  In the next section we describe a new SO(10) SUSY GUT
which satisfies all our criteria.  In fact it may contain the minimal symmetry breaking sector
consistent with the requirements of (1) obtaining the effective fermion mass operators of paper I, (2)
``naturalness," and (3) retaining only those states at energies below
$M_G$, which either have trivial SM charge or are contained in the minimal supersymmetric
standard model [MSSM].

\section{A Complete SO(10) SUSY GUT}
\subsection{The GUT symmetry breaking  and Higgs sectors}

For this model,
\begin{eqnarray}
W_{sym\, breaking}=& {1\over M} A_1' (A_1^3+{\cal S}_3 S A_1+{\cal S}_4 A_1 A_2) \\
& + A_2 (\psi \overline{\psi} + {\cal S}_1 \tilde A) + S \tilde{A}^2 \nn \\
& + S' (S {\cal S}_2+A_1 \tilde A) + {\cal S}_3 S'^2 \nn
\end{eqnarray}
where the fields \{$A_1,\; A_2,\; \tilde A, \; A'_1$\}, \{$S, \; S'$\}, $\psi, \; \bar\psi$,
\{${\cal S}_1, \cdots , {\cal S}_4$\} are in the 45, 54, 16, $\bar{16}$ and 1 representations,
respectively. [Note, traces and contractions of indices are implicit.]  The
supersymmetric minimization condition ${\partial W \over \partial A_1'}=0$ gives four discrete
choices  for the direction of the vev of $A_1$. We will assume that nature chooses the B~-~L direction. The second term gives $\tilde A$ a vev in the X direction, $A_2$ a vev in
the Y direction, and $\psi$ and $\overline\psi$ vevs in the right-handed neutrino-like direction
(see eqn. (\ref{eq:a1}) below). The third and fourth terms, and the ${\cal S}_4$ subterm of
the first term, were added to give mass to all non-MSSM fields which are not in a singlet
representation of the standard model gauge group. The last term was added in accordance
with our ``naturalness" criterion, namely that the theory should be the most general one
consistent with the symmetries. The term ${\cal S}_3 S'^2$ is consistent with the U(1) and R
symmetries of the symmetry breaking sector of the theory that will be discussed below, and we are
aware of no other additional symmetry that might exclude this term. Therefore the term must be
included.

The above vevs are given by
\begin{equation} \langle A_1 \rangle = a_1 \left(\begin{array}{ccccc} 1 &  & & &  \\
& 1 & & & \\ & & 1 & & \\ & & & 0 & \\ & & & & 0  \end{array}\right) \otimes
\eta  \equiv  a_1 {3 \over 2} (B - L)  \label{eq:a1} \end{equation}

$$ \langle A_2 \rangle = a_2 \left(\begin{array}{ccccc} 1 &  & & &  \\
& 1 & & & \\ & & 1 & & \\ & & & -3/2 & \\ & & & & -3/2  \end{array}\right)
\otimes \eta  \equiv a_2 {3 \over 2} Y  $$

$$ \langle \tilde{A} \rangle = \tilde{a} \left(\begin{array}{ccccc} 1 &  & &
&\\  & 1 & & & \\ & & 1 & & \\ & & & 1 & \\ & & & & 1  \end{array}\right)
\otimes \eta  \equiv  \tilde{a} {1 \over 2} X  $$

$$ \langle S \rangle = s \left(\begin{array}{ccccc} 1 &  & & &  \\
& 1 & & & \\ & & 1 & & \\ & & & -3/2 & \\ & & & & -3/2  \end{array}\right)
\otimes {\bf 1}   $$

$$ \langle \psi \rangle =  v \, \, |{\rm SU(5)
\, \, singlet} \rangle$$

$$\langle \overline{\psi} \rangle =  \overline{v} \, \, |{\rm SU(5)
\, \, singlet} \rangle$$
where $$ \eta = \left( \begin{array}{cc} 0 & -i \\ i & 0 \end{array} \right)
\; ,  {\bf 1} =  \left( \begin{array}{cc} 1 & 0 \\ 0 & 1 \end{array} \right)
\; .$$

The vacuum minimization conditions are explicitly
\begin{equation}
a_1^2=-s {\cal S}_3, \qquad {\cal S}_1 \tilde a = {1\over4}v \overline v 
\end{equation}
$$s {\cal S}_2+{2\over5}a_1 \tilde a=0, \qquad a_2 {\cal S}_1+2 s \tilde a=0$$
Using these equations, the set \{$a_1, a_2, \tilde a, v, \overline v, {\cal S}_4$\}
form a complete set of independent variables.

Two caveats --

$\bullet$ Note at tree level the GUT symmetry breaking vevs are undetermined since the potential
in these directions are both supersymmetric and flat.  We will not discuss the process for fixing
these vevs in this paper.  That analysis must necessarily include supersymmetry breaking effects
as well as supergravity and radiative corrections. 

$\bullet$ We do not describe the source of
supersymmetry breaking in this paper.  Soft SUSY  breaking operators are included at the GUT scale
and renormalized down to the weak scale when making any comparison with the low energy data.

The same doublet-triplet splitting mechanism that was used in paper II can be used in this 
model.  Accordingly, the Higgs sector of the Lagrangian is given by
\begin{equation} L_{Higgs}= [10_1 A_1 10_2 + {\cal S}_5 10^2_2 |_F + {1 \over M} [z^* 10_1^2 |_D \end{equation}
where the first term is $W_{Higgs}$ and the second generates a $\mu$ term when the F component
of the hidden sector field $z$ gets a vev of order $10^{10} GeV$.  We then obtain $\mu =
\langle F_z \rangle/M$.

\subsubsection{U(1) and R symmetries of $W_{sym\,breaking}$ and $\epsilon_3$}

$W_{sym\,breaking}$ has a [U(1)]$^4 \times$ R symmetry as is summarized in table
\ref{t:u1}. Up to arbitrary Yukawa coupling coefficients assumed to be of $O(1)$,
$W_{sym\,breaking}$ is the  most general superspace potential consistent with these symmetries.
Using $a_1, a_2, \tilde a, v, \overline v$, and ${\cal S}_4$ as independent variables, the only
invariant under a [U(1)]$^4 \times$R rotation of the vevs is $\zeta={a_1^4\over a_2^2 {\cal
S}_4^2}$.  After evaluating $\epsilon_3$ explicitly using equation (\ref{eq:grandexpression}) we
obtain

\begin{equation}
{\epsilon}_3={3 \tilde\alpha_G \over 10 \pi} \biggl\{ 2 \; \log{256\over 243} -
\log\abs{(1-25  \zeta)^4 \over (1-\zeta)}+2 \log \abs{{\tilde{M_t} \over M_G}} \biggr\}.
\end{equation}

Before we can check whether the experimental measurement of $\alpha_s(M_Z)$ in this model is
consistent with the non-observation of proton decay,  we must discuss the fermion mass sector of the theory.  The proton lifetime and branching ratios depend crucially on the couplings of the color triplet Higgses to fermions.  In a theory of fermion masses, these couplings are related to the Yukawa couplings of fermions to Higgs doublets, {\em but they are not identical}.  For example, the dimension 5 operators responsible for proton decay are given by
\begin{eqnarray} & {1 \over \tilde M_t} \;  {\bf Q} {1 \over 2} C_{QQ} {\bf Q} \; {\bf Q}  C_{QL} {\bf L} & \\ & {1 \over \tilde M_t} \;  {\bf \bar u} C_{ue} {\bf \bar e} \; {\bf \bar u} C_{ud} {\bf \bar d} & \nonumber \end{eqnarray}
where  $C_{QQ}, C_{QL}, C_{ue}$ and $C_{ud}$ are 3$\times$3 complex matrices.   The  matrices $C_{QQ}$ and $g_u$, the up quark Yukawa matrix, contain the same independent parameters but the SO(10) Clebschs are different.  An example is  presented in the next section.

\subsection{Fermion mass sector}

We take the flavor sector of our model to be that of model 4 in paper I.
Preliminary results of Blazek, Carena, Wagner and one of us (S.R.)\cite{bcrw} suggest that this
model provides the best fit to the low energy data.

The flavor sector is specified by a particular set of four operators $\{
O_{33}, O_{32}$, $O_{22}, O_{12} \}$.  Three of these operators --
$O_{33}, O_{32}$, and $O_{12}$ -- are uniquely specified by choosing model 4.
On the other hand there are 6 choices for operator $O_{22}$, labelled $(a, \cdots, f)$, as all 
give identical entries in the charged fermion Yukawa matrices.  As a result we can construct  6
possible models $4 (a, \cdots, f)$.

The four effective fermion mass operators for model $4c$ are given by
\begin{eqnarray}
    O_{33} = &  16_3\ 10_1 \ 16_3 & \label{eq:4coper}\\
     O_{23} = & 16_2 \ {A_2 \over {\tilde A}} \ 10_1 \ {A_1 \over {\tilde A}}   16_3  & 
\nonumber \\
    O_{22}  =  &  16_2 \ {{\tilde A} \over {\cal S}_M} \ 10_1 \ {A_1 \over
{\cal S}_M} \ 16_2 & \nonumber \\
    O_{12} = & 16_1 \left( {{\tilde A}\over {\cal S}_M}\right)^3 \ 10_1 \left(
{{\tilde A} \over {\cal S}_M} \right)^3 16_2 & \nonumber 
\end{eqnarray}

The superspace potential for the complete theory above the GUT scale which reproduces model 
$4c$  is given by (see fig. 1)
\begin{eqnarray}
 W_{fermion} = &    &     \\
 &16_3 10_1 16_3   +  {\bar \Psi}_1 A_1 16_3  + & {\bar \Psi}_1 {\tilde
A} \Psi_1  +  \Psi_1 10_1 \Psi_2   \nonumber \\
 & +  {\bar \Psi}_2 {\tilde A} \Psi_2  +  {\bar \Psi}_2 A_2 16_2  +  &
{\bar \Psi}_3  A_1 16_2  \nonumber  \\
 &   +  \Psi_3 10_1 \Psi_4   + & {\cal S}_M
\sum_{a=3}^9  ( {\bar \Psi}_a \Psi_a )  \nonumber \\
 & + {\bar \Psi}_4 {\tilde A} 16_2  +  {\bar \Psi}_5 {\tilde A} \Psi_4  + &
{\bar \Psi}_6 {\tilde A} \Psi_5  \nonumber  \\
 & +  \Psi_6 10_1 \Psi_7   +  {\bar \Psi}_7 {\tilde A} \Psi_8  + & {\bar\Psi}_8  {\tilde A} 
\Psi_9  +  {\bar \Psi}_9 {\tilde A} 16_1  \nonumber
\end{eqnarray}

 This superpotential is consistent with the symmetries discussed
previously with the addition  of one new U(1) given in table ~\ref{t:u1}. 
However it is not the most general fermion sector consistent with these symmetries.  In fact 
{\em one and
only one} new operator must be added
\begin{equation}
{\bar \Psi}_6 A_2 16_3  .
\end{equation}
It is easy to see that this operator leads to one new effective operator at the GUT scale 
when heavy states are integrated out (see fig. 2).  The new operator is
\begin{eqnarray}
 O_{13} = & 16_1 \left( {{\tilde A}\over {\cal S}_M}\right)^3 \ 10_1 \left(
{ A_2 \over {\cal S}_M} \right) 16_3 &  .\label{eq:13oper}
\end{eqnarray}
The complete model 4c is thus defined with this new operator and includes the operators in eqn. (\ref{eq:4coper}) and (\ref{eq:13oper}).  

A fit of model 4c to the low energy data\cite{bcrw} for certain ranges of soft SUSY breaking parameters agrees to better than  $1 \sigma$ for all observables.  A global $\chi^2$ analysis of models 4, 6 (a - f) (paper I) for
all regions of parameter space is now underway\cite{bcrw}.  It is already clear, however, that an additional operator such as the 13 operator in eqn. (\ref{eq:13oper}) is absolutely necessary to fit the data.  Whether this model fits the data better than any other remains to be seen.
For example, a different choice of 22 operator results in a different U(1) symmetry and thus by ``naturalness," a
distinct theory.  In particular, we have checked that for 
model 4b there are no new effective fermion mass operators
generated,  while for model 4a the new 13 operator 
\begin{eqnarray} O_{13} = & 16_1 \left( {{\tilde A}\over {\cal S}_M}\right)^3 \ 10_1 \left(
{\tilde A A_2 \over {\cal S}^2_M} \right) 16_3 & 
\end{eqnarray} is needed.

Finally, consider the matrices relevant for proton decay for the case of model 4c. In particular, the matrix $C_{QQ}$ is given by
\begin{eqnarray}  C_{QQ} & = \left(\begin{array}{ccc} 0 &  C & {1 \over 3} D e^{i\delta} \\
 C & -{1 \over 2} E e^{i \phi} & { 1 \over 3} B \\
{1 \over 3} D e^{i\delta} & { 1 \over 3} B & A \end{array} \right) &.
\end{eqnarray}
This should be compared with the up quark Yukawa matrix in the same model given by
\begin{eqnarray}  g_u & = \left(\begin{array}{ccc} 0 &  C & {1 \over 3} D e^{i\delta} \\
C &  0  & -{4 \over 3} B \\
-{4 \over 3} D e^{i\delta} & -{ 1 \over 3} B & A \end{array} \right) &.
\end{eqnarray}
where the 7 Yukawa parameters $A,\;B,\;C,\;D,\;E, \; \phi$ and $\delta$ are obtained by 
fitting to quark and lepton masses and the CKM mixing angles\cite{bcrw}.  Note that the
Clebschs in the two matrices differ by as much as a factor of 4.  These Clebschs affect
proton decay branching ratios. It is thus important to calculate the branching ratios in
models which are consistent with the observed fermion masses and mixing angles.

\subsubsection{Additional threshold corrections at $M_G$}
The dominant effective operators at the scale $M_G$ are obtained by integrating out states with 
mass greater than $M_G$.  Higher order corrections to these operators are also obtained.  These
typically lead to $O(10\%)$ corrections to the leading terms in the Yukawa matrices.  Of course
the terms in the Yukawa matrices will also receive corrections at one loop.  We have neglected
these corrections in the following analysis.

\subsubsection{Neutrino masses}
For completeness we include a minimal neutrino mass sector 
\begin{equation}
W_{neutrino} =  \bar{\psi} \; \sum_{i= 1}^3 \;16_i \; N_i
\end{equation}
where the states $N_i, \; i=1, \cdots, 3$ are SO(10) singlets.
This term has the effect of giving GUT scale Dirac masses to the right-handed neutrinos in the
16's.   Thus, the SM left-handed neutrinos are absolutely massless in this theory.  

If necessary,
left-handed neutrinos can be given masses as described in paper II\cite{hr}.  However, in order to
do so we must either break one linear combination of the U(1)s or introduce additional SO(10)
singlets.  In either case, we must then check for ``naturalness" and add any operator allowed by
the symmetries of the theory.

\subsection{Symmetries}
This theory has 5 global U(1) symmetries and a global continuous R symmetry.  The charges of
most of the states are given in table \ref{t:u1}.  The charges of the other states
can easily be derived.  We have checked our
theory for ``naturalness."  We find that only 3 additional operators need to be added to
the superpotential --$$\bar\psi_2 A'_1 \psi_1  {\cal S}_2, \;\; \bar\psi_2 A'_1 16_3  {\cal
S}_3, \;\; \bar\psi_5 A'_1 \psi_3  {\cal S}_3.$$ These three operators have no direct
effect on any observable properties since the vev of $A'_1$ vanishes.  With the inclusion
of these three operators the total superspace potential for model 4c is ``natural" (i.e. no
additional operators consistent with these symmetries are allowed) {\em for all possible
powers of the fields}.   In addition the theory has a matter reflection symmetry  (see
Dimopoulos and Georgi, ref.~\cite{DRW})  which forbids dimension 4 baryon or lepton
number violating operators.  

Some problems, however, remain to be solved in our model.  Given the states and symmetries, we find that $f_{\alpha \beta}$, the coefficient of the general
gauge kinetic term, is trivial in this model.  Thus we have not explicitly included the sector of
the theory which generates gaugino masses once SUSY is broken.
In addition, the U(1) symmetries of the theory are not sufficient to significantly constrain the
Kahler potential.  For example, terms such as $${ 1 \over M^2} \; \psi_4^* \;{\cal S}_M^*
\; \tilde A \; 16_2$$ are allowed which mix light generations with heavy states.  This term 
(and others like it) is allowed since it already
appears in the structure of the Feynman diagrams of fig. 1.
These off diagonal terms in the Kahler potential can affect fermion mass operators as well as
introduce flavor changing neutral current processes at low energies\cite{hkr}.     When
deriving the effective theory at $M_G$ we have implicitly assumed that the Kahler potential is
universal for all 16s in the theory.

\section{Results for $\alpha_s(M_Z)$ vs. Proton Decay in model 4c}

We now check whether the new model is consistent with $\alpha_s$ and the proton lifetime.
We have calculated the proton decay rate for the three dominant modes --- $K^+ \bar{\nu},$
$\pi^+ \bar{\nu},$ $ K^0 \mu^+$  where for neutrino modes we sum over the three neutrino
species.  We have included both gluino and chargino loops, as well as LLLL, LLRR and RRRR operators
in our analysis, where L(R) refers to left-handed (right-handed) fermion fields.  We used
values for the dimensionless Yukawa parameters at $M_G$ which give predictions for fermion
masses and mixing angles in excellent agreement with the data\cite{bcrw}.  The values for
soft SUSY breaking parameters are also consistent with electroweak symmetry breaking and the
experimental measurement of $b \rightarrow s + \gamma$.  We renormalized the dimensionless
(dimensionful) parameters  at two (one) loops to low energies in order to
make contact with the data. A detailed report on this work is in preparation\cite{lr}. For the
calculation presented here we take the effective Higgs triplet mass, $\tilde{M_t} =
a_1^2/{\cal S}_5 = 4 \times 10^{19} GeV$ with $a_1 = M_G \approx 2\times10^{16} GeV$ and
${\cal S}_5 = 10^{13} GeV$.  This corresponds to light Higgs doublets with mass  $10^{13}
GeV$.  

Is it natural to have such light Higgs doublets?  Are we populating the GUT desert?  To 
address this question we should compare the Higgs doublet mass with the spectrum of masses
for states in the symmetry breaking sector of the theory.  These in fact range from $10^{13}
- 10^{20} GeV$.  Thus we have taken the doublet mass to lie at the lower bound of this GUT
scale spectrum.  This seems to be the only natural criteria for setting an upper bound on
$\tilde{M_t}$. 

 The results for proton decay
are given in table \ref{t:pdecay} for two different values of soft SUSY breaking parameters 
and dimensionless couplings\cite{bcrw}.  For comparison we
also present the branching ratios obtained in a generic minimal SU(5) SUSY GUT from Hisano et
al.\cite{PDECAY}. The soft SUSY breaking parameters at the GUT scale are for (case A)
$M_{1/2} = 250, m_0 = 750, \mu = 64, m_{H_u} = 904, m_{H_d} = 1200 $ GeV and (case B) 
$M_{1/2} = 100, m_0 = 3000, \mu = 322, m_{H_u} = 4200, m_{H_d} = 3150 $ GeV with for both
cases, $A_0 =0$ and $\tan\beta = 53$.  The experimental lower bound on the proton lifetime
into the mode $K^+ \bar{\nu}$ is $10^{32}$ years\cite{kam}.  Thus these values are consistent
with the non-observation of proton decay to date.  However, these values are to be considered
as {\em upper limits} on the proton lifetime.  In particular,  $\tau_p\; Br^{-1}(p \rightarrow
K^+ \bar{\nu})$ scales as $({\tilde{M_t} \over 10^{19} GeV}{0.003 ~GeV^3\over \beta})^2$,
where $\beta$, with values in the accepted range $\beta = (0.003 - 0.03) GeV^3$\cite{beta},
is a measure of the matrix element of a 3 quark operator between the proton state and the
vacuum.  

Are these upper bounds on proton decay consistent with the experimental measurement of
$\alpha_s(M_Z)$?  We have evaluated the expression for $\epsilon_3$ including the additional
states with GUT scale masses contained in the fermion mass sector of the theory. These
typically shift the value of $\epsilon_3$ by a small amount in the positive direction. We find 
that for typical values of $a_1, a_2,$ and ${\cal S}_4$ around the GUT scale, we can
obtain  $\epsilon_3$  negative for $\tilde{M_t}$ of order $10^{19}$ GeV. For example, with $a_1 = 2 a_2
= 2 {\cal S}_4 = {\tilde a \over 3} = M_G $ we have $\zeta = 16$ and $\epsilon_3 \approx -0.030$,
which includes a positive contribution of $0.005$ from the fermion mass sector.

We have also checked that for these values of the parameters the gauge coupling satisfies
one loop perturbativity up to $M = 10^{18}$ GeV with $\alpha_G(M) = 0.39$.
Note that above the GUT scale we use the threshold boundary condition $$\alpha_G^{-1}(M_G) = 
\tilde\alpha_G^{-1} + \Delta_2(M_G)$$.

\section{Conclusion}  
We have presented a new complete SO(10) SUSY GUT.  This theory has several interesting features.  
The superspace potential is ``natural" to all orders in the fields.  It contains what may
possibly be the minimal GUT symmetry breaking sector necessary to obtain the desired adjoint vevs
consistent with (1) ``naturalness" and (2) fermion masses.  

We have also shown that one loop GUT scale threshold corrections are a significant constraint 
on the GUT symmetry breaking sector of the theory.  This constraint, for example, is
sufficient to rule out the model of paper II\cite{hr}.  The one loop threshold corrections
relate the prediction for $\alpha_s(M_Z)$ to the proton lifetime.  

 Finally we have calculated one loop GUT scale
threshold corrections to gauge couplings and  proton decay rates to different final states in
the new model.  The results are consistent with the low energy measurement of $\alpha_s(M_Z)$.  
The proton decay branching ratios provide a powerful test for theories of fermion masses. 
There is reasonable hope that new results from SuperKamiokande or Icarus could confirm or
rule out these theories.

It would be presumptious of us to conclude without discussing some of the open questions.  We 
have not discussed the origin of SUSY breaking nor how it feeds into the visible sector with
the one exception of the $\mu$ term which we have nominally considered.   We also do not
discuss what determines the GUT vevs.  At tree level, in the globally supersymmetric theory
considered here and neglecting SUSY breaking, these are flat directions of the potential. 
Finally, and perhaps most seriously, the symmetries discussed in this paper do not
significantly constrain  the Kahler potential.  Flavor mixing in the Kahler potential could
lead to dangerous flavor changing neutral current processes. We have {\em assumed} the trivial
universal Kahler potential in our analysis.

\acknowledgments
We would like to thank to Tom\'{a}\v{s} Bla\v{z}ek, Marcela Carena and Carlos Wagner for letting us use the results of work in progress on a general $\chi^2$ analysis of fermion masses.
This research was supported in part by the U.S. Department 
of Energy contract DE-ER-01545-640.

\newpage

\begin{center} {\bf APPENDIX 1} \end{center}

{\bf Proof of equation (\ref{eq:epsilon})}

 We define the GUT scale $M_G$ as the point where
$\tilde\alpha_G \equiv \alpha_1(M_G)=\alpha_2(M_G)$.  This means that
$\Delta_1(M_G)=\Delta_2(M_G)$.  We then define the relative shift in $\alpha_3(M_G)$ by
\begin{eqnarray} \epsilon_3 & \equiv & (\alpha_3(M_G)- \tilde\alpha_G)/\tilde\alpha_G \\
				& = & \alpha_3(M_G) \,(\Delta_3-\Delta_1 |_{M_G}) \nonumber 
\end{eqnarray}

 We then have
\begin{equation}
\Delta_3-\Delta_1|_{M_G} = {1 \over 2 \pi} 
(\sum_\gamma (b_3^\gamma-b_1^\gamma) \log \abs{{\det}' M_\gamma} - 
\sum_\gamma (b_3^\gamma-b_1^\gamma) n_\gamma \log M_G)
\end{equation}
where $n_\gamma=\hbox{the mass dimension of }{\det}' M_\gamma$ and ${\det}' M_\gamma$ is defined in eqn. (\ref{eq:effdet}), except for ${\det}' M_d$ where it will be convenient to define $\tilde M_d$ by
$\det \tilde M_d=M_G \,{\det} M'_d$, and ${\det}'M_d = {\det} \tilde M_d$.  This redefinition does not affect eqn. (\ref{eq:epsilon}).  Note, the matrix $\tilde M_d$ is defined such that $n_t = n_d$.

In addition,
$$\Delta_1 |_{M_G}=\Delta_2 |_{M_G}$$
which implies
\begin{equation}
\sum_\gamma (b_1^\gamma-b_2^\gamma) \log\abs{{\det}' M_\gamma} =  
(\sum_\gamma (b_1^\gamma-b_2^\gamma) n_\gamma) \log M_G
\end{equation}

Substituting for $\log M_G$ we obtain
\begin{eqnarray}
\epsilon_3 & \approx & {\tilde\alpha_G \over2\pi} \bigl\{ \sum_\gamma (b_3^\gamma-b_1^\gamma)
\log \abs{{\det}' M_\gamma} - {{\sum_\gamma (b_3^\gamma-b_1^\gamma) n_\gamma} \over
{\sum_\gamma (b_1^\gamma-b_2^\gamma) n_\gamma}} \sum_\gamma (b_1^\gamma-b_2^\gamma) \log
\abs{{\det}' M_\gamma} \bigr\} \nonumber\\ 
& = & {1 \over 2 \pi} {1 \over c_{12}} \sum_\gamma(b_1^\gamma
c_{23}+b_2^\gamma c_{31}+b_3^\gamma c_{12}) \log \abs{{\det}'  M_\gamma} \label{eq:epsint}
\end{eqnarray}
where $c_{ij}=\sum_\gamma (b_i^\gamma-b_j^\gamma) n_\gamma$. 
To evaluate the $c_{ij}$s, define $n_{54}$, $n_{45}$ and $n_{10}$ to be the number of 54, 45, and 10 representations in the theory, respectively, and
 $n_{16}$ and $n_{\overline{16}}$ to be the number of supermassive 16  and  $\overline{16}$ representations, respectively. For any SO(10) model built
with only 1, 10, 16, $\overline{16}$, 45, and 54 representations and one pair of light Higgs doublets, we have
$$\begin{array}{rcl}
n_{16} &=& n_{\overline{16}} \\
n_g = n_w &=& n_{45}+n_{54} \\
n_x &=& n_{45}+n_{54}-3 \\
n_u = n_e &=& n_{45}+n_{16}-3 \\
n_s = n_\sigma &=& n_{54} \\
n_q &=& n_{45}+n_{54}+n_{16}-3 \\
n_d = n_t &=& n_{10}+n_{16}
\end{array} $$
Evaluating $c_{23}$ explicitly
\def\nfi{n_{54}}
\def\nfo{n_{45}}
\def\nsi{n_{16}}
\def\nten{n_{10}}
\begin{equation}
\begin{array}{rcl}
c_{23}&=&(4 n_s+3 n_q+2 n_w+3 n_x+n_d)-(5 n_s+2 n_q+n_u+3 n_g+2 n_x+n_t) \\
&=& \{4 (\nfi)+3 (\nfi+\nfo+\nsi-3)+2(\nfi+\nfo)+\\
&& 3(\nfi+\nfo-3)+(\nten+\nsi) \\
&& - [5 (\nfi)+2(\nfi+\nfo+\nsi-3)+(\nfo+\nsi-3)+3(\nfi+\nfo)+\\
&&2(\nfi+\nfo-3)+(\nten+\nsi)]\} \\
&=& -3  
\end{array}  \label{eq:c23}
\end{equation}
Similarly
\begin{eqnarray}
c_{31}=9 \label{eq:c31} \\
c_{12}=-6 \nn 
\end{eqnarray} 
Thus, the $c_{ij}$s are completely independent of the number of fields in any theory built with
1s, 10s, 16s, $\overline{16}$s, 45s, and 54s.
Plugging eqns. (\ref{eq:c23}) and (\ref{eq:c31}) into eqn. (\ref{eq:epsint}), eqn. (\ref{eq:epsilon}) readily follows.

\begin{center} {\bf APPENDIX 2} \end{center}

{\bf U(1) symmetries and the dependence of $\epsilon_3$
on GUT scale vevs}

In this appendix we prove that the contribution to $\epsilon_3$ from the GUT symmetry
breaking sector is only a function of U(1) and R invariant products of powers of vevs.  The
proof relies on two facts: 
\begin{enumerate}
\item the effective determinants of mass matrices are holomorphic functions of the symmetry
breaking vevs, and
\item the effective determinants have simple phase rotations under U(1) and R
symmetry transformations.
\end{enumerate}
Note, since the effective determinants are independent of the conjugates of vevs, the U(1) and R invariance of $\epsilon_3$ is very restrictive.

We first discuss the case for mass matrices which do {\em not} include vector multiplets,
followed immediately by the case for mass matrices including vector multiplets.  Note, in the
first case the mass matrices themselves are, by construction, holomorphic functions of vevs. 
This is however not true for the latter case which is why it requires a separate discussion.

Consider a general
superspace potential $W(\Phi_1, \Phi_2, \ldots, \Phi_N)$ whose superfields  rotate under a
U(1) symmetry as $$\Phi_j \mathrel{{\mathop\to^\theta}} e^{i Q_j \theta}  \Phi_j$$ By
defining the shifted fields $\hat \Phi_i \equiv \Phi_i-\langle\Phi_i\rangle$ and expanding 
the superspace potential about $\langle\Phi\rangle$ we can find the fermion mass matrices.
\begin{equation} W(\hat\Phi_1+\langle\Phi_1\rangle,\ldots)= \ldots+\sum_\gamma \psi_\gamma
m_\gamma(\langle\Phi_1\rangle,\langle\Phi_2\rangle,\ldots) \psi_\gamma + \ldots
\end{equation} Now consider what would happen if the superfields received a different vacuum
expectation  value, $$\langle \Phi_j \rangle^{\hbox{\tiny new}} = e^{i Q_j \theta} \langle
\Phi_j \rangle^{\hbox{\tiny old}}$$ Under this change, the mass matrices would change.
\begin{equation} W(\hat\Phi_1+e^{i Q_1 \theta} \langle \Phi_1
\rangle,\ldots)=\ldots+\sum_\gamma \psi_\gamma  m_\gamma^\theta \psi_\gamma+\ldots
\end{equation} where $$m_\gamma^\theta \equiv m_\gamma(e^{i Q_1 \theta} \langle \Phi_1
\rangle, e^{i Q_2  \theta}
\langle  \Phi_2 \rangle, \ldots)$$
However, if we rotate the shifted fields $\hat\Phi_j$ by $e^{i Q_j \theta}$, the
superspace potential will be invariant under the combined rotation of $\hat\Phi$ and 
$\langle \Phi \rangle$.
\begin{equation}
\begin{array}{l}
W(e^{i Q_1 \theta} \hat\Phi_1+e^{i Q_1 \theta} \langle \Phi_1 \rangle, \ldots,
e^{i Q_N \theta} \hat\Phi_N+e^{i Q_N \theta} \langle \Phi_N \rangle) \\
\\
= \ldots+\sum_\gamma \psi_\gamma 
\pmatrix{e^{i Q_1 \theta} \cr
& e^{i Q_2 \theta} \cr
&& \ddots} 
m_\gamma^\theta 
\pmatrix{e^{i Q_1 \theta} \cr
& e^{i Q_2 \theta} \cr
&& \ddots} 
\psi_\gamma +\ldots \\
\\
= W(\hat\Phi_1+\langle \Phi_1 \rangle, \ldots, \hat\Phi_N+\langle \Phi_N \rangle) \\
= \ldots+\sum_\gamma \psi_\gamma \ m_\gamma \ \psi_\gamma+\ldots
\end{array}
\end{equation}
Therefore, $m_\gamma^\theta$ rotates in a very simple way.
\begin{equation}
m_\gamma^\theta=\pmatrix{e^{-i Q_1 \theta} \cr
& e^{-i Q_2 \theta} \cr
&& \ddots} m_\gamma \pmatrix{e^{-i Q_1 \theta} \cr
& e^{-i Q_2 \theta} \cr
&& \ddots}
\end{equation}
Thus,
\begin{equation}
\det m_\gamma^\theta=e^{-2 i \theta \sum Q} \det m_\gamma  \label{eq:u1charge}
\end{equation}
where the sum is over all fields that have columns in the mass matrix.

Similar arguments can show that under an R symmetry rotation, $\langle \Phi \rangle \to
e^{i Q \theta} \langle \Phi \rangle$, $m_\gamma^\theta$ is equal to 
\begin{equation}
e^{i Q_W \theta} \pmatrix{e^{-i Q_1 \theta} \cr
& e^{-i Q_2 \theta} \cr
&& \ddots} m_\gamma \pmatrix{e^{-i Q_1 \theta} \cr
& e^{-i Q_2 \theta} \cr
&& \ddots}
\end{equation}
where $Q_W$ is the charge of the superspace potential under the R symmetry. Therefore, 
\begin{equation}
\det m_\gamma^\theta= e^{i Q_W N \theta-2 i \theta \sum Q} \det m_\gamma \label{eq:rcharge}
\end{equation}
where $N=\dim m_\gamma$.

The situation is a bit more complicated for mass matrices which receive contributions from 
vector multiplets. The proof that the determinants and hence the effective determinants have simple phase rotations under the U(1) and R symmetry transformations can readily be extended to these mass matrices.  However, since the entries in the gaugino-chiral fermion mixing rows are actually the
complex conjugates of vevs, the determinants of these matrices are not holomorphic.  However, in the following we prove that the {\em effective determinants} of these
mass matrices, which include vector multiplets, {\em are} in fact holomorphic functions of the vevs.

Since the would-be Goldstone fermion states are perpendicular to the massive chiral fermion
states, any mass matrix containing gaugino-chiral fermion mixing can be written in the form
\begin{equation} \pmatrix{ 0 & x_1^*&x_2^*&x_3^*&\cdots&x_N^* \cr \overline x_1^*	&
c_{11}&c_{12}&c_{13}&\cdots&c_{1N} \cr \overline x_2^* & c_{21}&c_{12}&c_{23}&\cdots&c_{2N}
\cr \overline x_3^* & c_{31}&c_{12}&c_{33}&\cdots&c_{3N} \cr \vdots &\vdots&\vdots&\vdots&
\ddots & \vdots \cr \overline x_N^* & c_{N1}&c_{N2}&c_{N3}&\cdots&c_{NN} \cr} \end{equation}
with $\sum_j c_{ij} x_j=0 \hbox{ for all }i$, $\sum_i c_{ij} \bar{x}_i=0 \hbox{ for all }j,$ 
and the $c_{ij}$s, $x$s, and $\overline x$s are functions of the vevs but not of their
conjugates. The rows and columns of the mass matrix can be rearranged so that $x_N$ and
$\overline x_N$ are not zero. However, the determinant of this matrix can be reduced by
elementary row and column operations. Namely, by adding to the last column the second column
multiplied by $x_1\over x_N$ plus the third column multiplied by $x_2\over x_N$, and so
forth, the determinant of the mass matrix becomes $$\left| \matrix{ 0 &
x_1^*&x_2^*&\cdots&x_{N-1}^*&x_N^*+{x_1^* x_1\over x_N}+\ldots+{x_{N-1}^* x_{N-1}\over x_N}
\cr \overline x_1^*	&	c_{11}&c_{12}&\cdots&c_{1,N-1}&0 \cr
\overline x_2^* & c_{21}&c_{12}&\cdots&c_{2,N-1}&0 \cr
\overline x_3^* & c_{31}&c_{12}&\cdots&c_{3,N-1}&0 \cr
\vdots &\vdots&\vdots&\ddots & \vdots&\vdots \cr
\overline x_N^* & c_{N1}&c_{N2}&\cdots&c_{N,N-1}&0 \cr
} \right|$$
Doing the analogous operation on the rows, the determinant becomes
$$\left| \matrix{
0 & x_1^*&x_2^*&\cdots&x_{N-1}^*&{1\over x_N}\sum_k^N x_k^* x_k \cr
\overline x_1^*	&	c_{11}&c_{12}&\cdots&c_{1,N-1}&0 \cr
\overline x_2^* & c_{21}&c_{12}&\cdots&c_{2,N-1}&0 \cr
\vdots &\vdots&\vdots&\ddots&\vdots & \vdots \cr
\overline x_{N-1}^* & c_{N-1,1}&c_{N-1,2}&\cdots&c_{N-1,N-1}&0 \cr
{1\over\overline x_N}\sum_k^N \overline x_k^* \overline x_k & 0& 0& \cdots&0&0 \cr
} \right|$$
Thus, the determinant is equal to 
\begin{eqnarray}
{\sum_k^N x_k^* x_k \over x_N}{\sum_k^N \overline x_k^* \overline x_k \over \overline x_N}
\left| \matrix{
c_{11} & \cdots & c_{1,N-1} \cr
\vdots & \ddots & \vdots \cr
c_{N-1,1} & \cdots & c_{N-1,N-1}
} \right| \\
\nn \\
= (\sum_k^N x_k^* x_k) (\sum_k^N \overline x_k^* \overline x_k) f({\rm vevs}) \nn
\end{eqnarray}
where $f$ is a holomorphic function of the vevs. By setting $\overline v=v$,
$\overline x_i$ will equal $x_i \hbox{ for all }i$ and the mass of the vector multiplet is
${\sqrt{\sum_k^N x_k^* x_k}}$. Therefore, the determinant is equal to 
$M_{vector\,multiplet}^4$ times $f$ and the effective determinant is just the function $f$.
Note, $M_{vector\,multiplet}$ is thus always canceled from the denominator of the effective
determinant (eqn. \ref{eq:epsilon}) and no conjugates of vevs can appear in the effective
determinant. Thus, the effective determinants for mass matrices containing
gauginos are holomorphic functions of the vevs; just as those discussed earlier for mass
matrices which do not have gaugino-chiral fermion mixing entries.

These simple transformation properties of the mass matrices, under U(1) rotations of the vevs,
have significant consequences for the form of ${\epsilon}_3$.
Consider the following expression entering ${\epsilon}_3$ (see eqn.
\ref{eq:grandexpression}) --- \begin{equation}
{3\over 2}\log{{\det}'{M_g \over M_G}} -
  {3\over2}\log{{\det}'{M_w \over M_G}} +{33\over10}\log{{\det}'{M_s \over
M_G}} -{21\over10}\log{{\det}'{M_\sigma \over M_G}} \label{grandexpression} 
\end{equation} 
$$+{9\over10}\log{{\det}'{M_u \over M_G}}
+{3 \over 10}\log{{\det}'{M_e \over M_G}} -
  {6\over5}\log{{\det}'{M_q \over M_G}}$$
It is now easy to show that it is invariant under the U(1) and R symmetries.
Namely, the U(1) charge of the determinant of $M_w$ (eqn. \ref{eq:u1charge}) is equal to the 
charge of the determinant of $M_g$ which is equal to -2 times the sum of U(1) charges of all
fields in the 24 representation of SU(5). Therefore, the U(1) rotation of ${\det} \,M_g$ will
cancel the rotation of ${\det}\, M_w$ in expression, eqn. (\ref{grandexpression}). In addition,
we note that the U(1) charges of the effective determinants of \{$M_s$, $M_\sigma$\},
\{$M_u$, $M_e$\}, and  $M_q$  equal  -1 times the sum of U(1) charges of all fields in the
\{15 and $\overline{15}$\};  \{10 and $\overline{10}$\},  and \{10, 15, $\overline{10}$ and
$\overline{15}$\} representations of SU(5), respectively.  Therefore, the U(1) rotation of
${\det}' M_q$ is canceled by the rotations of the effective determinants of $M_u, M_e, M_s,$
and $M_\sigma$ in expression, eqn. (\ref{grandexpression}). Similar arguments show that the expression in
eqn. (\ref{grandexpression}) is invariant under an R symmetry rotation. Thus finally we
arrive at the conclusion that the expression in eqn. (\ref{grandexpression}) is invariant
under the U(1) and R symmetries of $W_{sym.\,breaking}$ for any superspace potential built
with 1, 10, 16, $\overline{16}$, 45, and 54 representations of SO(10). Moreover, since expression, eqn. (\ref{grandexpression}), is holomorphic, the
contribution of the GUT symmetry breaking sector to $\epsilon_3$ is only a  function of U(1) and R invariant products of powers of vevs.   

The same is not true for the contributions from
either the Higgs or fermion mass sectors.  This is because both the Higgs and fermion mass sectors contain massless states that must be projected out of the mass matrices before the effective determinants are taken.  After this projection, the determinants of the resulting mass matrices are no longer holomorphic functions of the vevs. Nevertheless for the Higgs
sector we can prove a similar but limited result, namely, that $g \equiv {{\det} M[t] \over {\det} M[d]} = g(\zeta_1, \cdots, \zeta_m)$;
i.e., $g$ is a function of U(1) invariants only.  By eqns. (\ref{eq:u1charge}) and
(\ref{eq:rcharge}), we see that ${\det} M[t]$ and ${\det} M[d]$ transform in the same way
under the U(1) and R symmetries; therefore the ratio is invariant.

%%%%%%%%%%%%%%%%%%%%%%

\protect
\begin{table}
\caption{SU(5) decomposition of SO(10) fields in our theory }.
\label{t:su5decomp}
\begin{eqnarray}
54 &\rightarrow &24+15+\overline{15} \nonumber \\
45 &\rightarrow &24+10+ \overline{10}+1 \nonumber \\
16 &\rightarrow&10+\overline5+1 \nonumber \\
10&\rightarrow&5+\overline5 \nonumber
\end{eqnarray}
\end{table}

\protect
\begin{table}
\caption{Notation used for states in different charge sectors}
\label{t:states}
$$\begin{array}{|c|c|c|}
\hline
{\rm SU(3)} \times {\rm SU(2)} \times {\rm U(1)\;  representation} & {\rm name} &
{\rm appears\; in\; SU(5)\; representation} \\ \hline
 (8,1,0)  &  g  & 24 \\
 (1,3,0)  &  w  & 24  \\
 (3,2, -{5 \over 3});\; (\overline 3,2, {5 \over 3})  &  x; \overline x  & 24 \\
 (3,1, {4 \over 3}); \;(\overline 3,1, -{4 \over 3}) &  u; \overline u  &  \overline{10} ;
10 \\  (1,1, -2); \;(1,1, 2) &  e; \overline e  &  \overline{10} ; 10 \\ 
 (3,2, {1 \over 3});\; (\overline 3,2, -{1 \over 3}) &  q; \overline q  & 10, 15;\;
 \overline{10} ,  \overline{15}  \\  (6,1,-{4 \over 3});\; (\overline 6,1,{4 \over 3}) &  s;
\overline s  & 15;  \overline{15}  \\  (1,3, 2);\; (1,3, -2) &  \sigma, \overline\sigma  &
15;  \overline{15}  \\  (3,1, -{2 \over 3});\; (\overline 3,1, {2 \over 3}) &  t; 
\overline t  & 5; \overline{5} \\  (1,2, 1); \;(1,2, -1) &  d; \overline d  & 5;
\overline{5} \\ \hline
\end{array}$$
\end{table}

\protect
\begin{table}
\caption{U(1) and R charges of the new model}
\label{t:u1}
$$\begin{array}{|c|l|c|l|}
\hline
\rm{field} & \hbox{U(1) charge} & \rm{field} & \hbox{U(1) charge} \\
\hline
A_1 & (1,0,0,0,0) & A_1' & (-3,0,0,0,0) \\
A_2 & (0,1,0,0,0) & \tilde A & (0,0,1,0,0) \\
S & (0,0,-2,0,0) & S' & (-1,0,-1,0,0) \\
\psi & (0,-1,0,1,0) & \bar{\psi} & (0,0,0,-1,0) \\
{\cal S}_1 & (0,-1,-1,0,0) &
{\cal S}_2 & (1,0,3,0,0) \\
{\cal S}_3 & (2,0,2,0,0) &
{\cal S}_4 & (2,-1,0,0,0) \\
{\cal S}_5 & (2,0,0,0,-4) &
16_3 & (0,0,0,0,1) \\
16_2 & (-1,-1,2,0,1) & 16_1 & (-2,-5,7,0,1)\\
10_1 & (0,0,0,0,-2) & 10_2 & (-1,0,0,0,2) \\
z & (0,0,0,0,-4) & N_3 & (0,0,0,1,-1) \\
N_2 & (1,1,-2,1,-1) & N_1 & (2,5,-7,1,-1) \\
\hline
\end{array}$$
\footnotesize All fields, except $A_1'$ and $z$, have R charge 1. $A_1'$ ($z$) has R charge
0 (2). The superpotential has R charge 3.  The first 4 charges listed above contribute to states in the GUT symmetry breaking sector.
\end{table}

\protect
\begin{table}
\caption{Proton decay results for model 4c}
\label{t:pdecay}
$$\begin{array}{|l|c|c|c|}
\hline
   & \hbox{A} &   \hbox{B} & \hbox{C}\\
\hline
\hbox{$\tau_p \;Br^{-1}(p \rightarrow K^+ \bar{\nu})/ (10^{32} yrs.)$} & 66 & 3000 & \\
\hbox{${\Gamma(p \rightarrow \pi^+ \bar{\nu}) \over \Gamma(p \rightarrow K^+ \bar{\nu})}$} & 0.28 & 1.18 &  0.5\\ 
\hbox{${\Gamma(p \rightarrow K^0 \mu^+) \over \Gamma(p \rightarrow K^+ \bar{\nu})}$} & 0.0040 & 0.0045 & 0.0007 \\ 
\hline
\end{array}$$
\footnotesize For $\tilde{M_t} = 10^{19} GeV, \beta = 0.003\; GeV^3$. Note, 
$\tau_p\; Br^{-1}(p \rightarrow K^+ \bar{\nu})$ scales as $({\tilde{M_t} \over 10^{19} GeV}{0.003~GeV^3\over \beta})^2$. The soft SUSY breaking
parameters for case A and B are described in the text.  In C, we show the results from Hisano et al.\cite{PDECAY} for
comparison.  Those results are from a generic minimal SU(5) SUSY GUT.
 \end{table}
\end{document}